\documentclass[11pt]{amsart}
\usepackage[parfill]{parskip}    
\usepackage{graphicx, txfonts}
\usepackage{epstopdf}
\usepackage{hyperref}
\usepackage{amsthm, amssymb, amsfonts, amsmath, enumerate, xy, mathrsfs}
\usepackage[margin=1in]{geometry}
\input xy
\xyoption{all}
\theoremstyle{plain}

\newtheorem*{question}{Questions}

\theoremstyle{definition}

\theoremstyle{remark}

\numberwithin{equation}{section}

\begin{document}

\title{A Project Based Approach to Statistics and Data Science}

\author{David White}
\address{Department of Mathematics and Computer Science \\ Denison University
\\ Granville, OH 43023}
\email{david.white@denison.edu}

\maketitle

\section*{Abstract}
In an increasingly data-driven world, facility with statistics is more important than ever for our students. At institutions without a statistician, it often falls to the mathematics faculty to teach statistics courses. This paper presents a model that a mathematician asked to teach statistics can follow. This model entails connecting with faculty from numerous departments on campus to develop a list of topics, building a repository of real-world datasets from these faculty, and creating projects where students interface with these datasets to write lab reports aimed at consumers of statistics in other disciplines. The end result is students who are well prepared for interdisciplinary research, who are accustomed to coping with the idiosyncrasies of real data, and who have sharpened their technical writing and speaking skills.

\section{Introduction} \label{sec:intro}

The ubiquity of computers and computational methods has provided massive amounts of data regarding the world we live in, and computers also provide a powerful way to analyze such data. Data science is an emerging field at the intersection of statistics, computer science, and disciplinary applications. Thus, mathematics departments can expect increasing demand for courses in statistics and data science, and increasing enrollments if such courses are offered. Similarly, individuals with PhDs in statistics and data science will be in short supply, as they will be in demand from both high paying industry jobs, and colleges and universities of all types. As a consequence, it is important that mathematicians develop the ability to teach applied statistics and data science courses. This paper will describe the process I went through, as a mathematician with no formal training in statistics, to develop a two-semester applied statistics sequence for our mathematics majors. 
The process I used to develop these courses involved reaching out to colleagues in other departments and leveraging their applied statistics experience. I will describe how I did this, so that my approach can be used at other institutions.

There are many different ways to teach statistics courses. The simplest division is between focusing on theory and focusing on applications. On the theory side, the traditional ``mathematical statistics'' course focuses on probability theory and the mathematical properties of statistical distributions. If any datasets appear in the course, they have usually been pre-cleaned and guaranteed to satisfy the hypotheses of the statistical models (e.g. the hypotheses of the Gauss-Markov Theorem \cite{wooldridge}). 
Real-world datasets effectively never look this nice, and students who have only taken a mathematical statistics course do not know what to do with dirty data, or data that does not satisfy the required hypotheses. Unfortunately, the mathematical statistics course just described is the only exposure to statistics that most mathematics faculty had as undergraduates, because curricular reform in statistics has only taken off relatively recently. This paper will focus on applied statistics courses, designed to follow the GAISE Guidelines \cite{gaise-report}, Amstat Guidelines \cite{ams-guidelines}, and ASA Curriculum Guidelines \cite{asa-guidelines}, where the goal is for students to learn how to work with real data, how to build and test statistical models for the data, how to check the conditions for these models, and what to do when the conditions fail. As has been pointed out by the MAA's CUPM Guidelines \cite{cupm} (Content Recommendation 3), mathematics majors programs should include such applied data analysis content. The CUPM guidelines do not recommend a mathematical statistics course for all mathematics majors, so mathematics faculty are left with a dilemma: we can either ignore the CUPM guidelines and teach a course similar to what we took as undergraduates, a course that does not work for the majority of students \cite{moore-math-stat}, or we can ``take the plunge" and learn how to teach a truly applied statistics course, even if we have no experience with applied statistics. By describing the process I went through to develop my applied statistics sequence, I hope to convince the reader that the latter option is achievable.

I teach at a small liberal arts college where class sizes are capped at 24. My department had a mathematical statistics course on the books, but the person who developed that course left the university ten years ago. Pure mathematicians without any formal background in statistics were teaching this course more-or-less in the way described in the previous paragraph. The course never had a large enrollment, and was usually populated entirely by mathematics majors, all of whom had at least taken linear algebra. My colleagues asked me to revamp the course, to make it truly applied, and to reduce the prerequisite to Calculus 1. The new course was to be called Applied Statistics. They also asked me to create a sequel to the course, Statistical Modeling. My position in the department involves teaching courses in both mathematics and computer science, so I decided to make computation a fundamental part of the course, following recommendations in \cite{ASA2}. The applied statistics sequence I developed had almost no overlap with my department's previous mathematical statistics course. I de-emphasized probability theory, and tried to give students just enough theory to appreciate the importance of the hypothesis behind the various statistical tests and models the course covered. The course revolved around weekly labs where students analyzed real-world datasets, using the statistical computing software R and an interface called RStudio, and then wrote lab reports summarizing these analyses for a general audience. The course culminated in a final project where students brought together all the tools they had learned to analyze a dataset of their choosing. This project acted as a kind of capstone, along the lines discussed in \cite{capstone}. I will discuss how I assessed the efficacy of my approach towards achieving the desired student learning outcomes in Section \ref{sec:methodology}.

To make space to teach students R, and to teach about the various real-world datasets, I cut out almost all discussion of deriving formulas for standard errors or properties of probability distributions. I did not show students complicated formulas unless it was necessary (e.g. to see how confidence intervals decrease in size as sample size increases). Students never had to compute regression coefficients or standard errors by hand. Instead, our software computed these quantities, and students learned how to interpret the output of the software, how to verify that the software was being used correctly (i.e. checking the hypotheses of the model), and how to check that what they were observing was a real effect and not statistical noise (e.g. via cross-validation). This is not to say I taught a ``cook book" course: the course remained as rigorous as the mathematical statistics course, but the rigor was shifted away from algebraic manipulations (rendered unnecessary by the software) to carefully thinking about potential bias in the datasets, how to write code to clean the data, how to transform the data to satisfy the hypotheses the statistical models required, and how to write technical lab reports. Students completed several labs over the course of the semester, culminating in a final project. The collection of labs, the instructions I gave students for writing the labs, and the rubrics I used to grade the labs and final project are all hosted on the course webpage. For a mathematics department just beginning to teach statistics, I believe applications should be emphasized to provide students with a mental framework for understanding as many data situations as possible. An upside of this approach is that the prerequisites can be reduced, so that the applied statistics sequence can be made available to students from across campus. In addition, our present enrollments in the applied statistics sequence are much higher than our enrollments in mathematical statistics, so students are ``voting with their feet" that they prefer the applied approach. 


My process for developing the applied statistics sequence revolved around outreach to colleagues in partner disciplines. I leveraged the applied statistical expertise of these colleagues to determine what statistics content would be most appropriate for a year-long applied statistics sequence. Research papers written by these colleagues provide excellent material to create projects, as the specifics of each dataset have already been cataloged, as the conclusions students should reach are already known, and as the faculty member can serve as a project supporter if the student should have questions. In addition, such projects naturally set up students to do research with faculty members who contribute their data and analyses. In order to make this course repeatable, and to share resources with others on campus who teach statistics courses (e.g. research methods in social science, biostatistics, experimental physics, etc), I created a data repository hosted on the university server, that all contributing faculty had access to. The main section of this paper, Section \ref{sec:outreach}, describes the process I used for contacting colleagues in other departments, soliciting suggestions for content (especially for the second course in applied statistics), and building the data repository. I also give examples of labs I created using these datasets. 

Based on these discussions with partner disciplines, I designed the applied statistics sequence to cover content that working data scientists would need. A discussion of the topics for these two courses is provided in Section \ref{sec:applied-topics}. A selection of projects to support these topics, based on real-world datasets, is provided in Section \ref{sec:projects}. Section \ref{sec:methodology} provides a discussion of how I assessed the resulting courses. Lastly, the Appendix contains specific details about the implementation of these courses, how I structured the day-to-day classes to maximize the educational impact of the projects, and details about the course content. This Appendix is designed to help mathematicians develop an applied statistics course, if they have never taught one before.

\section{Outreach to Partner Disciplines} \label{sec:outreach}

At a small liberal arts college, ``partner disciplines'' often means the entire campus. While attempting to design the applied statistics sequence, I was fortunate to have access to helpful colleagues in a wide variety of departments. It was easy enough to find which colleagues in other departments used quantitative methods in their research. These colleagues were often the same ones who taught the ``research methods'' courses in their departments. Mathematically, these courses are equivalent to the lowest level statistics course in the mathematics curriculum, i.e. our service course for non-majors. When I learned I would be developing the statistics sequence, I reached out to statistically oriented faculty in psychology, neuroscience, biology, chemistry, physics, geoscience, economics, political science, education, sociology, and classics (computational linguistics). I also received datasets from our athletics department, from our institutional research staff (anonymized data on students from previous years), and from our investment office. Lastly, I received materials on machine learning from a computer science professor. All of these materials were put into a data repository, discussed in Section \ref{subsec:data-repository}. I found all these colleagues were willing to meet with me.

\begin{question} Here are questions I asked when meeting colleagues in other departments.
\begin{enumerate}
\item What statistical topics do you cover in your teaching?
\item What topics do you wish you could cover if you had more time?
\item What topics do you often find yourself teaching your summer research students to equip them to do research in your field?
\item What topics do you often need for your research, or that you remember from statistics you took in graduate school?
\item What are some commonly used sources of data in your field?
\item I'm in the process of creating a repository full of datasets, metadata, projects based on the data, and teaching materials. All contributors get access to the whole repository. Would you be willing to contribute the data from your research, or teaching materials from your course?
\end{enumerate}
\end{question}

Question (1) gave me a sense of which topics were most important to users of statistics, and hence in the first course, Applied Statistics. It also gave me a wide range of examples to use when covering those topics, and gave me valuable information about which courses on campus to allow as prerequisites for the Statistical Modeling course, in case students who didn't take the Applied Statistics course wanted to get in. Question (6) was a great way to get materials to flesh out the course, e.g. Excel spreadsheets and handouts working out examples. Questions (5) and (6) gave me access to a huge range of trustworthy data. I used some of this data in the projects I created, and also made it available to students for their final projects. Even colleagues who did not teach statistics in their curriculum (e.g. in chemistry, education, and classics) were able to give me access to datasets and online repositories. 

Questions (2)-(4) were geared more towards the Statistical Modeling course. Through these conversations I learned about the need for principal component analysis, non-parametric statistics, time series analysis, and the generalized linear model (especially Poisson regression) in our curriculum. These conversations also served as a valuable way for me to learn about these topics and to get references I could share with my students. For example, colleagues in physics and geoscience highlighted the importance of modeling measurement error, since their data comes from physical devices, and precision is a key concern. Based on an example from the experimental physics course, I developed a short unit on measurement error for Statistical Modeling. As another example, colleagues in economics, political science, and sociology highlighted the importance of more advanced linear regression techniques for when the conditions of regression (e.g. homoscedasticity, no autocorrelation, non-normally distributed errors) are not met. I devoted a large chunk of Statistical Modeling to such situations, and used the examples my colleagues cited as in-class examples and labs. Finally, colleagues in geoscience, biology, and psychology all agreed that principal component analysis is the most common tool in their research, and one they consistently teach to summer research students. Now that this topic is contained in my Statistical Modeling course, and now that the prerequisites have been arranged to allow entry for non-math majors, these professors plan to send future summer research students to take Statistical Modeling. 

These discussions with colleagues led to several pleasant side-effects. First, everyone I met with agreed to answer student questions if students selected their dataset for use with a final project. This meant that I did not need to become an expert on every dataset in the data repository. Secondly, I was able to act as an intermediary between different departments, to raise the overall level of statistical knowledge at the university. For example, I taught a colleague in chemistry about Q-Q plots, a tool for testing if a dataset is normal that I learned from a colleague in psychology. Similarly, I taught a colleague in neuroscience about some non-parametric methods I learned from a colleague in geoscience, to get around a lack of normality in his dataset. Thirdly, once faculty saw the wealth of material in the data repository, several expressed interest in setting up a reading group to discuss how best to teach statistical concepts across the curriculum. In the 2016-2017 school year, I worked with the director of our Center for Teaching and Learning to run such a reading group, with a regular attendance of more than 12 faculty. I created a syllabus of statistical pedagogy papers, and associated discussion questions. I later learned that various faculty groups over the past twenty years had called for such a gathering of faculty from diverse disciplines, to improve quantitative offerings across campus. 
I hope to write a paper in the near future discussing this experience and making the reading group materials public.
 
The last side effect relates to the new Data Analytics (DA) major at my university. This major was proposed in 2015, and adopted in 2016. The faculty in the reading group have been strongly supportive of the new major and have created applied tracks through their majors for these data analytics students. In the spring of 2017, Data Analytics 101 was offered for the first time, similar in spirit to Applied Statistics (using R, de-emphasizing theory, and analyzing real-world datasets from the data repository) developed and team-taught by a a psychologist, a political scientist, a biologist, and a mathematician. All these faculty were members of the reading group and contributors to the Data Repository. These faculty created four large-scale projects for DA 101, three of which used datasets from the Data Repository. In the near future, these materials will be made public along with the materials from my statistics sequence, and I will host a link at \href{http://personal.denison.edu/~whiteda/index}{my webpage} \cite{webpage}. The statistics sequence will serve DA majors, following the curricular recommendations of \cite{pcmi}.

\subsection{Data Repository} \label{subsec:data-repository}

Most universities have an internal server system where each faculty member can request storage space. As I gathered datasets and teaching materials from colleagues, I put them into a folder on our university server so that everyone teaching statistics could use the materials and could contribute to them. At a campus without a statistician, it seemed wise to leverage the abilities of all statistically-oriented faculty on campus. In this way, each of us could draw examples from a variety of fields, each of us could use labs developed by the others, and each of us could store the data we planned to use in a safe place in case it ceased to be available at the original source.

When I asked my colleagues Questions (3) and (4), they suggested statistical topics based on their research experiences. With a follow-up question, I was able to obtain the datasets where they needed these statistical methods. As a result, I had real data I could use to demonstrate each topic I wanted to teach, and to create projects for students. I found that students enjoyed replicating results published by their professors and by other students who did summer research. Additionally, my colleagues often had a deep understanding of the datasets they gave me, and together we could create metadata about the datasets. In this way, anyone using the data (either future faculty or students) would know what was in the data, what results to expect from students working on the data, and any issues with the cleanliness of the data. In some fields, this type of metadata is called a \textit{codebook}. I decided early on not to clean the data for my students, since the GAISE Guidelines are clear that students should learn to work with real data, which is often messy. However, I still wanted to know the ways in which the data was messy, e.g. outliers, missing data, saved in a bad file format, etc., so that I could walk the students through the exploratory data analysis and cleaning stage. 

In some fields the culture is to keep datasets private and only publish the findings from the data. The data repository allowed colleagues to make their data available for teaching purposes and only on our campus, so that the data would not find its way into the hands of a competing research group. I think the majority of my colleagues had some reticence about sharing data, but the friendly nature of our liberal arts college convinced them. Originally, I think they would have resisted making the datasets public, even for teaching purposes. However, there is currently a push on my campus to make the data repository (and associated teaching materials) publicly available, as part of my university's new major in Data Analytics. This new major received a great deal of support from the faculty I contacted in other departments, and most have agreed to allow their datasets to be made public in the coming months.

\subsection{Sample applications}

As a result of these conversations with partner disciplines, I created a list of datasets my students could draw from when choosing their final projects, and I created several labs, discussed in Section \ref{subsec:labs-repository}. An abbreviated list of datasets is provided below, to help faculty who are new to teaching statistics, but want to use real data. A complete list, with links to data sources, can be found on the course webpage for Statistical Modeling \cite{stat-modeling-webpage}. Most of the data for these projects can be found online, e.g. UCI data archive, ICPSR archive, Kaggle, Data Science Central, deeplearning.net, and KDnuggets.com.

\begin{itemize}
\item Sociology - city of Chicago data, American Time Use Survey, police data on traffic stops, Chinese Census, UK government data, Bureau of Labor Statistics, general social survey. 
\item Data Mining - from Google, Facebook, Twitter, Uber, Netflix, Wikipedia, Pandora, Pew Research Center, bike-share programs, traffic data.
\item Biology - genomics, ecological forecasting, healthcare data (NESARC, ADHEALTH), NIST.
\item Astronomy - Radio jet data, FITS image data.
\item Physics - protein folding, human movement data, data from the small Hadron collider.
\item Geoscience - ice core data for global temperatures, Paleobiology Database.
\item Neuroscience - field of vision, connectomics, classifying personality types via relevant factors.
\item Chemistry - energy in reactions, pharmacology data.
\item Psychology - PsycINFO repository, data on brain disorders and genetic correlations.
\item Political science - polarization in Congress, religious congregations, voting patterns.
\item Economics - Conference Board, World Bank, Yahoo Finance, salary data.
\item Computer science - social networks, data on passwords.
\item Education - NAEP database, CUPP report, Common Core data, Department of Education.
\item History - trans-Atlantic slave trade database, CIA Factbook, Global Terrorism Database.
\item Sports - baseball (MLB), football (NFL), soccer (FIFA), and basketball (NBA).
\item Computational linguistics - authorship verification, natural language processing, Google Ngram.
\end{itemize}

I provided a list much like this to prospective students at course registration time (to pique their interest), at the start of the semester (to motivate them), and when I handed out the prompt for the final project. About 50\% of the students chose from one of the datasets on the list. The others found their own datasets online. When a dataset from online is suitable for a final project, I add it to the data repository and use the student final project to create the metadata. In this way, the data repository is continually updated for future classes, whenever students find more usable data sources.

\subsection{Labs based on these datasets} \label{subsec:labs-repository}

As we will see in Section \ref{sec:projects} (and the Appendix), the labs in my statistics sequence were a mixture of labs drawn from statistics courses at other universities, projects taken from textbooks, and labs of my own design. In this section, we focus on labs created as a result of contacts with partner disciplines. 

An early homework assignment in Applied Statistics (the first course in the sequence) involves studying several misleading graphics used in the news media (that I compiled with the aid of professors in psychology and political science). For each graphic, students write a short descriptions of what the graphic represents, why it is misleading, and what human biases might cause a desire to mislead. Students are then directed to find another example of a misleading graphic. This assignment reinforces early content about data visualization, helps students to develop skepticism, and demonstrates the real-world utility of studying statistics. This homework takes place before students have learned R, so there is no actual data analysis in the lab.

Another pair of early assignments in Applied Statistics, while students are learning R, leads students through a data analysis of polling data. This unit is particularly popular in election years. I created this unit based on discussions with a political science professor, who pointed me to relevant datasets, methodology from Nate Silver's blog, and published sources on a few subtler points. In the first polling assignment, students analyze different graphics and write about which are most useful. Then, students conduct a rudimentary meta analysis, whose purpose is to reveal radical differences between polling aggregation website. Students write about bias, and how to model trustworthiness of a poll, then compute a simple confidence interval. In the second polling assignment, students work through increasingly sophisticated models for forecasting election results, beginning with simple linear regression, regression with interaction, and weighted regression. Students write about what makes forecasting difficult, and students must carefully list the assumption they are making for their forecast to be valid (e.g. assuming that no large events occur between now and election day to change public sentiment, assuming the trend in poll data retains the same shape at future moments in time, etc.). Students then read Nate Silver's methodology and critique it, based on what they learned.

A late RStudio lab in Applied Statistics, once students are already comfortable with R and with writing lab reports, instructs them to comb through the World Bank dataset to identify a question that they are interested in, then attempt to answer this question using multiple linear regression. Students will find that the conditions for regression are rarely satisfied: transformations to linearity are often required, scatterplots often display heteroscedasticity, and the data is spread across several years of time (leading to a likely violation of the autocorrelation condition). I also give the students an expository reading by an econometrician suggesting there is reason to believe that the World Bank data exhibits severe bias, since countries report their data knowing full well that the amount of funding they receive from the World Bank will be based on this data. In the end, a good lab report conducts an analysis along the lines of those we have covered in the course, with several caveats that the results should be viewed with some skepticism due to the failure of so many regression conditions. This lab can be frustrating for students, but often motivates them to continue on to Statistical Modeling (where we learn how to cope with autocorrelation, and where we learn more powerful tools of coping with heteroscedasticity). This lab also leads nicely into the final project, by giving students lots of flexibility, but also holding them accountable for exercising caution throughout their analysis. Some students even choose World Bank data for their final projects, and try to learn some of the content from Statistical Modeling to fix the issues identified in this lab.

An early RStudio lab in Statistical Modeling involves conducting an exploratory data analysis of human field of vision data gathered at my university by a professor of neuroscience and his students over several years. My students re-create the figures (histograms and boxplots) from a published paper, and then improve on the analysis. First, students recreate the two-sample t-tests and p-values in the published paper. Then, students are directed to use what they have learned from our in-class examples, about checking the conditions of a test before carrying it out, and using tests appropriate to the question at hand. Ideally, students will use Q-Q plots to discover that the normality conditions are not satisfied, then use non-parametric tests for a difference in means to see if this failure affects the main results of the paper (thankfully, it does not). Lastly, strong students will realize that ANOVA should be used for a question involving multiple comparisons, and will carry out a non-parametric version of ANOVA, the Kruskal-Wallis test, to minimize the chance of a Type I error.

A midsemester RStudio lab in Statistical Modeling analyzes a dataset about various species of birds, that I obtained from a colleague in biology. Students have flexibility in choosing the response variable, but are instructed to build the best multivariate regression model possible. This dataset is rife with missing data and impossible data. The dataset also requires transformations of data types to properly read it into R. Students must clean the data while building the model, because different choices of explanatory variables will have different missing data values to remove. The lab comes after a lengthy in-class discussion of p-hacking, so that students will realize the danger of trying numerous models and only reporting the one with the best $R^2$. For this reason, the lab requires students to use cross-validation, and to write in the Methods section about whether the way in which data is missing could lead to bias. In a more advanced class, this would be a good dataset to demonstrate multiple imputation and other methods for filling in missing data, or testing if the way the data is missing is random or not.

Lastly, several students chose datasets from the Data Repository for their final projects. Many of these projects could be adapted to future labs. For example, one student analyzed an internal university dataset with anonymized student data on merit-based financial aid, major, GPA, gender, race, and ranking by the admissions office upon acceptance. The project realized that admissions ranking and merit-based financial aid do not predict for GPA, raising serious questions about the efficacy of current administrative practice. Another student analyzed a dataset on automobile insurance from an alumnus who works in the area. This student presented her analysis during interviews and is now an actuary. Another student carried out an analysis on human movement data provided by a professor in the physics department, and later became her research assistant. Lastly, as part of a reading course with another professor in my department, a student used the tools from my course to analyze a criminology data provided by an alumnus (now part of the data repository), and later won an internship in his company to continue her analysis over the summer.

\section{Applied Statistics Topics} \label{sec:applied-topics}

This section discusses the content I emphasized in the two-semester statistics sequence I developed. My topic selection was informed by discussions with colleagues in partner disciplines, about what topics came up most often in their research, and by discussions with statistics professors at several liberal arts colleges. 

\subsection{Emphasizing real world applications}

The goal of the two-semester statistics sequence I developed, following the CUPM Guidelines \cite{cupm}, was to provide students with applied data analysis skills. For this reason, I structured the statistics sequence around weekly labs where students analyzed real-world datasets, as suggested by the GAISE Guidelines \cite{gaise-report}, culminating in a final project where students analyzed a dataset of their choosing (in 2016, this was called a ``Semester Long Project'' to encourage students to start it early). As a byproduct of this approach, students became familiar with datasets in numerous fields, and in particular with certain datasets used by my colleagues in their research. Several students went on to carry out research with those faculty members. Furthermore, when students applied for jobs, they had a portfolio of data analyses to show their skills. Many students reported that this helped them get a job, and that what they learned in the statistics sequence helps them in their jobs. The lab reports and final presentations also provide students experience with interpretation of their results in context, communication of results to non-technical audiences, and technical writing skills. On a daily basis, active learning techniques are used (see \cite{freeman} for the efficacy of this approach), and students are given hands-on opportunities to work with data.

While the idea of teaching via applied labs is not new in statistics (\cite{discovery-projects, halvorsen-projects, nolan-explorations, nolan-speed, nolan-book, beyond-normal, Wild and Pfannkuch}), this style of teaching is less popular in mathematics. When mathematicians teach statistics, theory is often emphasized over applications, and pre-cleaned data (usually, the data that comes with the textbook) is often used instead of real-world data. Emphasizing real-world data, messy data, and exploratory data analysis is much more useful for our students, and my experience suggests improved student retention when content is delivered in a hands-on, applied way, rather than in a way that emphasizes theory. By de-emphasizing theory, the course is free to include a wide variety of models, as well as an emphasis on the consequences of failures of model conditions. In particular, the course features a heavy discussion of randomization-based inference, a computer-reliant non-parametric method that is becoming increasingly important in the data analysis world. More details can be found in the Appendix.

I found the Isostat email listserv \cite{isostat} an extremely valuable resource. This listserv is populated by statisticians discussing statistics pedagogy. Many listserv members are at small liberal arts colleges. People on the listserv were willing to share textbook recommendations, projects they had used, and exam questions. They also helpfully explained concepts to me when I was first learning the material in these courses. I strongly encourage any mathematician assigned to teach statistics to consult this email list. It is free to join. For more about teaching statistics at liberal arts colleges, see \cite{stats-lib-arts, stats-lib-arts2}.

\subsection{First Course in Statistics}

Leveraging the mathematical maturity guaranteed by a calculus prerequisite, the Applied Statistics course I developed contains all the usual topics that a first course in statistics generally covers, plus a bit more: students learn how to visualize data, how to explore data, how to fit multivariate linear models to data (including ANOVA models), correlation vs. causation, basic tools from probability theory (taught via simulation methods as much as possible), sampling distributions, common statistical distributions (e.g. normal distribution), how to compute confidence intervals, and how to do inference regarding data and regarding regression models. This content usually takes about 80\% of the course, leaving roughly 2-3 weeks to cover advanced topics such as multi-way ANOVA, rank-based non-parametric tests, or logistic regression (or to simply slow down the first part of the course). The flexibility also allows for a deeper focus on R at the beginning of the semester, if a particular group of students needs it. 

A detailed schedule, created with help from professors of research methods courses in other disciplines, is available on the course webpage \cite{white-242-2016}. Briefly: the course begins with an overview of the statistical programming language R, daily homework from a website called DataCamp helps students build familiarity with R, and class sessions proceed in an interactive way, as I use a prepared RStudio workbook to analyze a dataset related to the night's reading, following suggestions from the class. As the students develop proficiency in R and basic statistical modeling techniques, via the OpenIntro labs \cite{open-intro} in DataCamp, we transition into spending more class time with students working in RStudio, and I begin to assign highly structured labs in RStudio (rather than DataCamp) following the model developed by Wagaman \cite{wagaman}, including several labs based on the datasets in the data repository. These labs feature increasingly dirty data, and increasingly more freedom for students to decide how to carry out their analysis (i.e. less and less structure in the skeleton RStudio file I provide to students). The labs culminate in the semester-long project. 

As a liberal arts professor, I've found that a first course in statistics can be an excellent place to teach students about skepticism and fallacies. For the unit on data visualization, the class covered numerous examples of media and politicians using such visualizations to mislead. For the unit on correlation, we discussed numerous logical fallacies (led by ``correlation does not imply causation'', of course, and supplemented by examples from the website \href{http://www.tylervigen.com/spurious-correlations}{http://www.tylervigen.com/spurious-correlations} \cite{spurious}). Lastly, when covering probability theory, we covered Bayes' Theorem and discussed applications to medical testing. The basic idea is that, if students ever get a positive test for some terrible disease, their probability of actually having the disease is lower than they might expect, and thus they should consider getting a re-test. The course devotes a substantial amount of time throughout to the dangers of ``p-hacking,'' where a researcher runs many tests until finding something interesting. Students read case studies and news articles, and write short responses to identify if whether or not a study should be viewed with suspicion and why. These sorts of real-world applications of the concepts of the course keep students engaged. I also presented this section of the course to the faculty reading group, and many faculty in other departments expressed interest in including a similar module in their own courses.

I have taught the Applied Statistics course twice. After switching from a mathematical/theoretical approach to an applied approach, enrollments boomed, so that now we offer multiple sections each semester, rather than one section per year. Additionally,  as word spread of the applied nature of the course, the proportion of non-math majors taking the class has drastically increased. Members of the faculty reading group, and contributors to the data repository, now encourage their students to take this class, and some have even sat in on the class themselves. The average GPA has not changed with the switch from my colleagues' mathematical statistics course to my Applied Statistics course, and students report on evaluations that the class is very challenging. Thus, the increased enrollment suggests students really prefer the applied focus of the course, even if it means learning R and statistics simultaneously.

\subsection{Second Course in Statistics}

There are many options for a second course in statistics (\cite{ams-guidelines, 2nd-design, cobb-2nd-mathy, wagaman}). The most common focuses on regression - multivariate linear regression, ANOVA, and logistic regression - with the punchline that ``everything is regression'' via the generalized linear model. This is the second course I chose to teach, and I used the Stat2 book \cite{stat2}. This book is extremely easy to read, and only requires that students have exposure to basic statistics and a bit of R. Hence, the prerequisites can be kept very low. Just like my Applied Statistics course, I emphasized real-world projects, building up to a final project. Each project contained a written report, in which students summarized their findings to an audience I gave them, e.g. to a senator, to a dean, to a CEO, etc. I devoted particular attention to building models, choosing between different models, the conditions required of the various models, and what remains true when these conditions fail to be met. A daily course plan including readings, weekly projects, daily R practice in DataCamp, and videos showing students how to program in R, can be found on the course webpage \cite{stat-modeling-webpage}: \href{http://personal.denison.edu/~whiteda/math401spring2016.html}{http://personal.denison.edu/$\sim$whiteda/math401spring2016.html}. Many of these materials were not developed by me. I found them with help from the isostat listserv \cite{isostat} and thus was able to focus my attention on creating the new labs discussed in Section \ref{subsec:labs-repository}, and included in the data repository. More details on the day-to-day structure of this course can be found in the Appendix.

One benefit of statistics over mathematics is the ability for students to take such a wide variety of classes after completing their first course, as opposed to the sequential nature of a mathematics major. At the end of the semester, all of my non-graduating students (and two who graduated but still lived in the area) requested more statistics, so I offered a seminar in the fall of 2016 on Bayesian statistics and survival analysis. Then, in the spring of 2017, I offered a seminar on data mining and time series analysis for the same students, joined by others from my Applied Statistics course from the fall of 2016. In the future, I hope to turn these seminars into real upper level electives, now that sufficient interest in statistics has been established.

\section{Data-driven projects} \label{sec:projects}

In this section, I describe projects created based on my conversations with partner disciplines, datasets and repositories shared by my colleagues, and teaching materials I received from statisticians via the isostat listserv. More details can be found in the Appendix.

\subsection{Applied Statistics Course} \label{subsec:242-projects}

My goal for Applied Statistics was that students would be capable of conducting a self-driven analysis of a dataset of their choosing. I wanted students to be able to detect issues of bias in a dataset, to clean a given dataset, to choose the correct procedures to analyze the dataset, and to write a report summarizing the results of their findings to a lay readership. Following the principles of backwards design, I decided that the course should culminate in a final project where students do precisely these tasks. I then developed labs to build students up to the final project, and I structured quizzes and exams to emphasize the skills needed to complete the final project (as discussed in Section \ref{sec:applied-topics}). I began the course with basic exercises in R, to supplement the unit on data visualization. I quickly transitioned into the Open Intro labs \cite{open-intro}, and in some weeks I was able to assign two such labs, leveraging the fact that my students all had more mathematical maturity thanks to the calculus prerequisite. Working through these labs, and the basic exercises in R, occupied the first half of the course (the first 7 weeks). During this time, students were learning more about RStudio in class, and at times even asked if they could use RStudio instead of DataCamp. Hence, in the second half of the semester, students were ready for the switch to RStudio.

At this time, students are given access to the data repository, and instructed to find a dataset either from the repository or from the internet. Students then write a two-page proposal for their final project. The proposal summarizes the dataset, identifies a response variable to study, suggests how the various topics from the course will be used in the analysis, and reflects on potential bias in the data. The prompt for this writing assignment is hosted at \cite{white-242-2016}. Students are not allowed to gather novel data, because at this point we have not discussed experimental design or issues related to the Institutional Review Board (IRB). 

For the second half of the semester, students complete weekly projects in RStudio, starting with highly structured projects very similar to those in DataCamp, and culminating with free form data analyses. These projects are based on those of Wagaman \cite{wagaman}, but some use datasets from the data repository, as detailed in Section \ref{sec:outreach}. During all labs in the latter half of the course, students are directed to make steady progress in their final projects. Each week, they are supposed to carry out tests and build models for their own dataset, based on
the content we are learning. In the final two weeks of the course, students are free to write their final papers, turning all the analyses they have done into a coherent story.  In the last week of the semester, students give presentations of their results, and hand in their final papers. In the past, student projects have replicated results from professors at my university (re-analyzing datasets from the data repository), have extended analyses from the published papers connected to the datasets in the data repository, and have conducted entirely new analyses. Once, a student project found a mistake in a published paper, and was able to conduct a correct analysis using the tools from the course (thankfully, yielding the same result as the published paper). 
On several occasions, I have felt student final projects were of publishable quality, and this is even more true of projects in Statistical Modeling. I am currently working with four different students to get their final papers ready to submit to undergraduate research journals.

\subsection{Statistical Modeling Course} \label{subsec:401-projects}

The learning goals for the Statistical Modeling course were similar to those for Applied Statistics: I wanted students to be able to carry out a detailed final project using the tools from the course. Hence, the Statistical Modeling course was also project-driven, now with both a midsemester project and a final project. I de-emphasized exams and quizzes to add weight to the projects, and each project had both a written part and an R part. In the written part, students wrote an Introduction, Results and Conclusions, and Methods section. The introduction was always written to a non-technical audience. The results section needed to contain just the punchlines and the two best graphics produced. The methods section would talk about potential biases, any decisions the students had to make about outliers, justification of the conditions for tests and models, and potential impact of their decisions to the final results. The written part was required to remain below 3 pages. Separately, the R part would walk students through an analysis, so that they could better understand the week's topic and how to implement it in R, and then would give them the dataset to analyze for their project. Students could include as many supporting tests and graphics as they wanted in this document (viewed as an appendix to the main paper). Some of the materials for these labs were adapted from \cite{kuiper-book}, some from Amy Wagaman of Amherst College, and some from colleagues in other departments (replicating their research papers, as described in Section \ref{subsec:labs-repository}). In the future, I will also design labs based on the midsemester and final projects of these students.

Early projects were extremely structured, so that students really only needed to interpret the output of code I gave them, or write basic commands of their own modifying that code. Later projects were not structured at all, instead just giving students space to write R code and conclusions. In addition to these projects, students completed midsemester and final projects on topics of their choosing. I required that the midsemester project involve an in-depth data analysis using both multivariate linear regression and logistic regression. The final project could be another data-driven project (required to use ANOVA, principal component analysis, or time series analysis), a theory project (e.g. for students interested in graduate school), or an R programming project. Several of these projects were excellent, and I encouraged students to turn them into publishable papers. For example, one data-driven project \cite{trevor-genome} models which of five learning disabilities patients had based on their genome. One theory project \cite{tybl} is a comprehensive overview of the field of spatial econometrics, where correlations based on geographic location are taken into account. One programming project \cite{trevor-R-code} conducts an exhaustive search over all exponents $(p,q)$ to attempt to make $y^p$ and $x^q$ linearly related. This code is now available on GitHub and can be imported into RStudio.

\section{Assessing the efficacy of this approach} \label{sec:methodology}

As discussed in Section \ref{sec:intro} and \ref{sec:applied-topics}, when I began teaching statistics, my department did not have any applied statistics courses. Now, we have the two semester sequence described in this paper, the new course DA 101 mentioned in Section \ref{sec:outreach}, and material for several more applied electives from the seminars I ran in 2016 and 2017 (see Section \ref{sec:applied-topics}). Enrollments in statistics increased in each of the past three years, leading to an additional section of Applied Statistics being offered in the fall of 2017. Noticeably, enrollments from non-math majors increased drastically.

On the formal end-of-semester evaluations, students are asked to rate how challenging they found the course, whether their interest increased, and whether their knowledge increased (among other questions). In Applied Statistics in 2016, 85\% of students reported that the course was challenging, 80\% reported that their interest increased, and 85\% reported that their knowledge increased. In Statistical Modeling, 100\% reported that the course was challenging, 84\% reported that their interest increased, and 100\% reported that their knowledge increased. I view these results as justification that these courses are not watered-down mathematics courses. They are challenging and thought-provoking in different ways than traditional mathematics courses, but lead to more interest from students, and higher enrollments.

I asked students to fill out an optional supplemental evaluation at the end of each semester. In Applied Statistics, there was a 100\% response rate, and in Statistical Modeling it was 92\%. In Applied Statistics, 85\% reported that they would like to take another statistics course. All reported that they believed the course content will be important in their careers. In Statistical Modeling, all students reported that they wish they had taken statistics earlier in their college careers. All reported that they believed the course content will be important in their careers. 82\% reported that they would take more statistics after Statistical Modeling, if it were offered. In surveys from both courses, 63\% reported that they would have considered graduate work in statistics (this percentage is significantly higher than the percentage of students my university sends to graduate school in any department). 

In terms of my learning objectives for the course, I can report that I am consistently blown away by the high quality of student final projects. I greatly enjoy watching students grow from their early days of being completely unable to use R, up through proficiency in the labs, and into mastery of R, data analysis, and technical writing for the projects. Several projects have been of publishable quality, and I'm working with four different students to get their final papers ready to submit to undergraduate research journals. 

Lastly, the development of the two-semester applied statistics sequence, in consultation with faculty from other departments and designed around the data repository, has had several beneficial consequences around campus. Since most departments who have contributed to the data repository already require calculus, the calculus prerequisite for Applied Statistics is no obstacle for their students. Faculty in physics, biology, chemistry, neuroscience, and economics have all encouraged their students to take Applied Statistics, and the current enrollment composition reflects this. As discussed in Section \ref{sec:outreach}, one positive consequence (that I hope to explore in a future paper) was the formation of a faculty reading group, discussing commonalities across the various quantitative methods courses on campus, sharing teaching materials, and updating course content accordingly. Multiple departments are now using R, are now teaching randomization-based inference, and are now emphasizing the importance of reproducible analyses, and the dangers of p-hacking. The faculty who shared data sets with the data repository have been happy to serve as project supporters when my students choose to analyze their data. When student analyses have gone beyond what the faculty member originally did, it has on several occasions led to the student conducting research with the faculty who contributed the dataset. Now more and more faculty are contributing to the data repository, and students from the new data analytics major will also be able to analyze these datasets, as part of their required self-driven data analysis projects in the major.

\section{Conclusion}

In this paper, I describe my experiences as a pure mathematician tasked with creating a two semester statistics sequence. I achieved this goal by reaching out to partner disciplines with a series of questions designed to help me create courses that could best serve their majors. I received course materials and datasets from these colleagues and housed them in a Data Repository. I also convinced these colleagues to serve as problem supporters for projects related to the datasets they shared. I synthesized the materials I received from colleagues with materials given to me by statisticians on the Isostat listserv \cite{isostat}, and added several labs and homework exercises of my own design (Section \ref{subsec:labs-repository}). 

This process resulted in two applied courses, built around daily real-world examples, weekly labs, and a final project. In the end, students learned how to cope with dirty data, how to wrangle data in R, how to build statistical models and run statistical tests, how to use statistics to debunk or support claims about the world, and how to summarize their findings to a non-technical audience. Students strengthened their liberal arts skills, such as critical thinking skills, developing an interdisciplinary perspective on the world, considering the ethics of certain types of data and experimental design, and healthy skepticism when presented with statistical arguments. Through their individual projects, students also learned how to teach themselves how to do things in R, what to do when their data situation didn't match anything we'd learned, and how to check their conclusions against what they know about the world in order to have confidence that they'd done the analysis correctly. 

After developing these courses, I shared my course materials with the same colleagues in partner disciplines. As a result, these colleagues requested a reading group to discuss statistical pedagogy across the campus, and several of them went on to develop a data analysis course using materials from the Data Repository. I have made my course materials publicly available on the course webpage, and am in the process of making the Data Repository and all associated labs freely and publicly available through the university's new Data Analytics major.

I believe the model outlined in this paper can be replicated at other liberal arts colleges. I recommend starting with a well-respected colleague in another department, who is friendly towards the mathematics department. In my case, I began with a neuroscience professor who was also chair of the faculty at the time. Once this other professor is on-board with contributing to the data repository, I recommend reaching out to other friendly professors until a critical mass is formed. After enough professors have joined, the incentives are higher for reluctant professors - after all, every professor who joined the data repository received access to the course materials of all others who had joined. If relationships between the mathematics department and other departments are strained, then some mending of fences may be required in order for the approach discussed in this paper to work, but new faculty members might still have a chance to make inroads with other departments.

When I began this process, I did not know if statistics courses had a canonical selection of topics - the research methods courses across campus had some overlap but in general emphasized different statistical modeling frameworks (e.g. econometrics focused on regression, psychology focused on experimental design, and physics focused on measurement error). In the Appendix, I have provided details about the content I settled on, based on conversations with partner disciplines and other statisticians. I believe this content, and the emphasis on labs and final projects, provides enough depth so that students can conduct meaningful statistical analysis, and is broad enough so that nothing essential is missing. However, it is worth noting that other selections of content are possible, and might be preferable based on the types of statistical modeling conducted by colleagues in other departments. Unlike the mathematics major, statistics students are often not expected to have all seen the same core content. It is more important that they know how to look for issues of bias, how to wrangle real-world data, and how to learn new modeling frameworks as needed. Thus, a different selection of topics might be perfectly fine. In any event, the community of statistics professors on the Isostat listserv \cite{isostat} is extremely friendly, and I strongly recommend contacting the listserv often as any new statistics course is being developed.

Lastly, I wish to include a word about developing applied statistics courses within traditional mathematics departments. I am fortunate to have a group of senior colleagues who strongly encourage junior faculty to experiment, and who value mathematical modeling. In a department less open to change, and with no statistician on the faculty, it may be dangerous for a junior faculty member to attempt such a drastic course overhaul. In this situation, I recommend reaching out to the statistics community via Isostat \cite{isostat} and trying to invite senior statistics professors to visit (e.g. as part of a colloquium). When the mathematics department sees that statisticians are teaching statistics in the applied way described in this paper, they may be more open to allowing a junior faculty member to mimic this approach. If possible, having a statistician involved in an external departmental review would also provide an impetus to the mathematics department to be more open to applied statistics courses. Finally, there is a copious literature and curricular recommendations about teaching applied statistics courses, that I have tried to make accessible to the reader via the bibliography (notably, \cite{gaise-report, ASA2, ams-guidelines, pcmi, asa-guidelines}, and \cite{cupm} all contain curricular recommendations for including more applied statistics), and sharing some of these sources with senior faculty might make them more open to applied statistics content within the mathematics major.

\appendix

\section{Further details on the Applied Statistics Course}

As discussed in Section \ref{sec:applied-topics}, the philosophy of the Applied Statistics course is to de-emphasize theory, in order to make space for non-trivial analyses of real-world datasets using R, culminating in a final project where each student analyzes a dataset of his/her own choosing. In this appendix, I provide more details for how to carry out such a course. This appendix will be most useful to a mathematician who has never taught statistics before. For the benefit of such a reader, I have at times attempted to explain points I found somewhat subtle, but which might be trivial to a professional statistician. I have also emphasized details about the choice of textbook, prerequisite, and how I kept the grading manageable, among other topics. For readers who already teach an applied statistics course, this appendix can be skipped.

Before diving into the nuts and bolts of the courses I created, I wish to include one more word about the decision to emphasize applications over theory, for the benefit of a reader concerned about the preparation of students bound for graduate school. For institutions where the majority of math majors are bound for graduate school, it might make sense to emphasize theory over applications, but for a mathematics department just beginning to teach advanced statistics, I believe applications should be emphasized to provide students with the analytical framework for understanding as many data situations as possible. Once firm interest in statistics has been developed (as discussed in Sections \ref{sec:applied-topics} and \ref{sec:methodology}, this has already occurred at my institution), an upper level mathematical statistics course can be introduced to serve students bound for graduate school in statistics or data science. Before that time, such students can achieve depth in theory via mathematics courses such as real analysis. 

\subsection{Topics}

\subsubsection{On the use of R}

Based on formal evaluations and on informal Google Forms I ask students in the statistics courses to fill out, I have learned that students find learning R the hardest part of the course. It is worth noting that there are many choices of statistical computing language (e.g. SPSS, SAS, Stata, JMP, StatCrunch \cite{statcrunch}), but generally speaking, students find all of them difficult \cite{biehler, R-relative, R-versus}. I chose R because it's what most of my colleagues in other departments were using in their research, even though none of them taught using R. It is worth noting that students find just as much difficulty with other statistical computing languages \cite{R-relative, R-versus}. For more on the decision of which statistical software to use, see \cite{biehler}. 
For most students, this is the first time they have needed to run code or cope with error messages. It is tempting, even for statistics professors, to put the burden on the students to teach themselves R outside of class. However, as discussed in \cite{nolan-computing-curr}, this sends students the wrong message about what is and is not important, and also leads to students developing bad habits and wrong conceptual understandings. Similarly, I have found that, in order for students to feel comfortable with R, in-class R time is essential. After reading about R commands, and working through several examples as part of a group, students need to actually type out lines of code and run them. While R is painful for students at the beginning of the semester, students are very pleased with both R and their skills in R by the end of the semester. In order to get student buy-in at the start, when R seems very hard to them, the first assignment of the year\footnote{This idea is due to Rick Cleary, of Babson University} requires students to gather data on how much money an R programmer makes and then compute summary statistics for this data in R.

I begin the course with a very simple R interface, and then introduce RStudio to highlight R's graphical capabilities (this goes well with the unit on data visualization). Once students have RStudio working, I assign labs and homework from a website called DataCamp. 
This website walks students through basic commands in a game-like environment, where students earn ``experience points" for correct answers. Students are instructed to cap their time spent on homework at 30 minutes per night, and homework is correspondingly weighted little in determining the final grade. Homework is meant to help students identify gaps in their knowledge, so that they can ask questions in class or in office hours. Labs are worth substantially more than homework. The first several labs in my Applied Statistics course are drawn from Open Intro \cite{open-intro}. Even though the Open Intro labs are designed for a service course, rather than a course with a calculus prerequisite, when it comes to R the calculus prerequisite seems to matter little. Using the Open Intro labs makes grading laughably easy, as students can simply submit a screenshot demonstrating completion of the lab (with the experience points, so that I know they actually typed out the commands). I find that I can usually assign two Open Intro labs in a single week, and students are not overwhelmed. During class, I continue to use RStudio, and I continue to assign in-class RStudio exercises to help students transfer what they are learning in DataCamp into the RStudio environment. Using R also makes the probability theory in the course easy to cover; R is excellent for computing probabilities, demonstrating the laws of probability theory via simulation, and running Monte Carlo simulations. Students appreciate that, with this tool, they can always figure out the answer to the sorts of probability theory questions they saw in high school, without working through a theory-based derivation of the probability in question.

On a day-to-day basis, I use an interactive lecture style to work through an example data analysis in RStudio. The example is already typed out, but I use a guided discussion to reveal what should be done in each step. If students suggest approaches different from the approach I prepared, I type the code in real time to carry out the suggested student approach, and then compare it to the one I had prepared. Then, students work in groups to answer some easy conceptual questions and then to do an analysis of their own in RStudio. This is facilitated by the use of a laptop cart that my university has, though most students prefer to bring their own laptops to class each day. During this activity, I walk around to check that students completed their reading notes (conceptual questions on the reading from the previous night) and to answer questions. Daily activities for students are in the Data Repository, but were drawn heavily from Kaplan's book \cite{kaplan}. 

In order to make sure students are working through the example R code from class, I give several quizzes where students are allowed to use the example R code. Each quiz question requires students to figure out which in-class topic is related, then tweak the code from that topic in order to answer the question. These quizzes reflect that my course is more about recognition, modification, and synthesis of content, rather than memorization of content. Exams contained a part where students could use R, and a non-R part with only conceptual questions. In this way, my examinations I shift focus away from rote memorization, and towards concepts, decision-making on real-world datasets, and knowing where to look to find R code to modify (i.e. recognizing which dataset from class a dataset on the exam is most similar to). I believe these cognitive tasks are more important than memorization, for the modern world. However, students are not allowed to use Google or any resources other than the R code provided to the class over the course of the semester.

After a few weeks, once I feel students are getting comfortable with R, I switch from DataCamp to highly structured RStudio labs, modeled off those used by Wagaman \cite{wagaman}.
In these labs, students are shown examples, asked to compile code and interpret the results, and asked to write a few lines of code themselves mimicking the examples they have seen. The resulting HTML and PDF files produced by RStudio are quite lengthy, because they contain all the text I wrote to guide the students, plus the text students wrote to answer my questions, the code students wrote, and the output produced by that code. To streamline the grading, I put the word ``ANSWER'' everywhere where students are supposed to put an answer. That way, I can grade the HTML or PDF file by using the search function for the word ``ANSWER.'' After a few highly structured labs, I begin to lessen the structure and give students more flexibility in how to conduct their analyses. I also give them dirtier and dirtier data. By the time students start working on their final projects, they have learned several techniques for handling dirty data and coping with uncertainty in a self-driven analysis. All of my R materials are contained within the Data Repository, so will soon be freely available online at \cite{webpage}. After my success with R, the biology department switched from JMP to R, the new data analytics major uses R throughout, and the psychology department is in the process of switching from SPSS to R. 

Even though coping with dirty data is painful, students always report that it is extremely valuable and they wish we could spend even more time on it. I introduce dirty data early, during the unit on visualizations, where we can see empty histograms (e.g. if R is interpreting integers as strings), impossible data (e.g. a large number of 0s in blood pressure measurement, if the person collecting the data denoted by 0 a lack of a measurement), and gaps in data (e.g. in regression, if data was only kept for certain blocks of years). A taxonomy of dirty data types is provided in \cite{taxonomy-dirty}, and a discussion of common scenarios students find themselves unprepared for is provided in \cite{preprocessing}. I try to cover a variety of types of dirtiness, based on these sources. Once students are comfortable spotting oddities in the data, I give them some tools for coping with the types of dirtiness we see, e.g. having R drop `NA' entries, using subsetting to manually drop rows with certain characteristics (e.g. impossible ages), and using general commands to read in data with a designated type rather than with the default type. Via readings on the reproducibility crisis in psychology and other data analysis fields (see \cite{reproducibility} for a nice survey), I impress upon students the importance of doing their cleaning in R, rather than using Excel. As with any content, watching me clean data is much less useful than having students do it themselves. So, after our first RStudio lab, I start giving students dirtier and dirtier data. At the beginning, it's impossible data, sometimes even with a codebook describing the scheme used to mark missing data. Then, in the military spending lab \footnote{adapted from a conversation with Shonda Kuiper in 2016}, students deal with trying to join two datasets that name countries slightly differently (one dataset comes from the CIA factbook, the other is internal to R's ggplot package and is used for drawing maps). Lastly, labs on the avian and World Bank datasets (described in Section \ref{subsec:labs-repository}), as well as student final projects, get students to a certain level of depth with cleaning data. The hope is that the collection of techniques they have learned will suffice for many real-world situations after graduation, and that exposure to such a wide variety of techniques will enable students to come up with their own as needed.

The first time I taught the Applied Statistics course, I used a new book by Akritas \cite{akritas}, because it contained lots of R code alongside the mathematics. In hindsight, this book still focused more on mathematical derivations than I wanted, and led to an overly mathy treatment \cite{applied-stats-webpage}. The second time \cite{white-242-2016}, I used Danny Kaplan's book \cite{kaplan}, and this strengthened the applied data focus of the course. Kaplan's book \cite{kaplan} also comes with reading questions, which I find a nice way to hold students accountable for doing the reading before class. This enables me to spend class time working through applied data analyses in R, or giving students conceptual questions designed to provide a high-level understanding of the mathematics behind the statistics. The hardest part for students, other than R, is the logic of inference and type I vs type II errors. For these units, I created supplemental handouts to support the treatment in the books above, and I always build in extra time to the schedule in case students are still having difficulty with these topics. These handouts are available in the Data Repository.

Kaplan's book \cite{kaplan} is fantastic for its conceptual emphasis, for teaching students to be read news stories that use statistics, and for real-world applications. As discussed in Section \ref{sec:applied-topics}, I find a first course in statistics an excellent place to teach skepticism. In addition to the examples discussed in Section \ref{sec:applied-topics}, we discuss medical testing at a higher level in the unit on ANOVA. When testing for a number of potential health problems, a doctor must decide whether to use multiple t-tests or ANOVA. The latter keeps a bound on the probability of a Type I error, i.e. on the probability of rejecting the null hypothesis when you should have failed to reject. If the null hypothesis is true, then this probability is just the shaded tail probability representing the rejection region, i.e. it's the level of significance, $\alpha$, of the test. If you do multiple t-tests then you're increasing the size of the overall rejection region. Often, this can be corrected by shrinking the size of $\alpha$ for each individual test (e.g. using a Bonferroni correction). This simple fact, that if you do multiple tests you're making it more likely that one will be significant, has real applications for students interpreting the results of a blood panel, or of a full body scan. It also connects nicely to a discussion of p-hacking, and of the reproducibility crisis in several fields (where published research cannot be reproduced). In this part of the course, I emphasize technical writing, I insist that students detail their methodology in a section of each lab report, and I teach students to make their data and analysis publicly available. 

\subsubsection{Randomization-based inference}

Other than the use of R and the labs, the most important difference between my course and a standard mathematical statistics course is my emphasis on randomization-based inference and bootstrap confidence intervals. The reason for emphasizing randomization-based inference is that they are non-parametric, meaning that they work even if the dataset is not normally distributed (or distributed according to another of the classical parametric distributions). As many datasets in the real world (and many in my class) are not normally distributed, non-parametric methods are critical for applied data analysts. An additional benefit of teaching with bootstrap methods is that one can delay probability distributions until later in the semester, and can de-emphasize the sampling distribution, one of the more conceptually challenging topics for students. For more about the pedagogical advantages of teaching via randomization-based inference, see \cite{cobb-ptolemaic,lock-primus, combating-simulation, tintle-rand-based, accessible-inference, accessible-bootstrap, wood}. 

The idea of randomization-based inference is to lean on modern computing power rather than mathematical approximations using the classical distributions. Rather than assuming the sample statistic follows one of these distributions, we instead assume that our sample is representative of the population (formally, that the population ``looks like'' a bunch of copies of the sample), and then use a Monte Carlo simulation to compute the p-value as (number of successes) divided by (number of trials). For example, if given paired data $(x_1,y_1),\dots, (x_n,y_n)$, we can test the hypothesis that the correlation $\rho$ is zero as follows. First, write down the sample correlation $r$ of this dataset (one could also make minor modifications to the discussion that follows to consider a difference of means $\overline{x} - \overline{y}$). If the true correlation were zero, then the pairing didn't matter. So we use our dataset to generate thousands of other datasets by randomly pairing $x_i$ values to $y_j$ values (i.e. breaking the given pairings). If the null hypothesis is true, then the original pairing did not matter, because the $x$'s and $y$'s are unrelated. Thus, under our assumption that the sample is representative of the population, we can pretend that these new datasets are drawn from the population, rather than created from the sample. With thousands of datasets, we can make an empirical sampling distribution, now called the \textit{bootstrap distribution}. To do so, for each dataset (i.e. each random pairing), we write down the correlation obtained, and we gather all the correlations together in a dotplot. 

The mean of the bootstrap distribution is close to zero, since the distribution was created by assuming the null hypothesis and breaking the given pairing. The standard deviation of the bootstrap distribution will be based on how much variability there was in our original dataset. It approximates the standard error of the sampling distribution. To obtain the p-value, we simply count the number of bootstrap samples whose correlation was larger than our original sample correlation $r$. This count is the number of successes in our Monte Carlo simulation. For example, if we created 3000 bootstrap samples (from 3000 random pairings), and if 36 had a correlation as large as our original sample correlation $r$, then the \textit{bootstrap p-value} of a 1-tailed test would be $36/3000 = .012$. Obviously, different randomization-based tests can yield different p-values, but asymptotically they converge. Empirical studies have shown that, if the dataset really is distributed according to a classical distribution, this bootstrap p-value is usually close to the p-value one would obtain via parametric methods. The mathematical theory justifying bootstrapping is detailed in \cite{hesterberg-bootstrap}. The original idea of the bootstrap goes back to Ronald Fisher, one of the giants of statistics, but, due to a lack of computing power, he introduced the $t$-test to approximate the result of the randomization-based test described above \cite[Section 4.4]{hesterberg-bootstrap}.

For a lower level statistics course, without a calculus pre-requisite, \cite{lock5} provides an approach entirely focused around randomization-based inference. This approach comes with software called StatKey that allows students to visualize this process. I made brief use of StatKey when first introducing randomization-based inference, but because my course revolved around labs in R, I quickly gave students the tools to conduct these analyses in R (these tools are also contained in Kaplan's book \cite{kaplan}). My Applied Statistics course was the first course on campus to teach using randomization-based inference, but now the new data analytics major also follows this model and teaches randomization-based inference before parametric inference, as suggested in \cite{lock-primus}. I have also found that starting the second course in the sequence with a week of randomization-based inference provides a useful review for students who took Applied Statistics, provides new content for students who entered Statistical Modeling from one of the research methods courses in another department, and highlights the applied data analysis philosophy of the course from the start.

\subsection{Projects}

Recall from Section \ref{sec:projects}, that an explicit goal of Applied Statistics was to prepare students to conduct self-driven data analysis, assessed via their final projects. At the start of the course, students had no background with either statistics or programming, so I wanted early projects to be highly scaffolded, unambiguous, and focused on how to conduct simple statistical procedures with R. As discussed in Section \ref{sec:applied-topics}, I used the Open Intro labs \cite{open-intro}. Once students were comfortable with R, I introduced highly structured RStudio labs focused on how to carry out the analyses we were learning on real datasets. Later in the course, building up to their final projects, students learned to clean data and make tough decisions with no clear right answer. Students greatly enjoyed being able to work through their early projects in this structured way, and they were ready for the harder and more ambiguous labs when those labs came. By that time, students had already completed their ``final project proposal'' detailing their plans for the final project, so students knew that they would need to be making hard decisions. 

The Open Intro labs \cite{open-intro} I used are summarized in the following list, along with the real world applications of each. The order is the order I used these labs, which differs from their given order on DataCamp.

\begin{enumerate}
\item Numerical summaries of data using R, and an analysis of eruptions of Old Faithful.
\item Visualizing data, and exploratory data analysis on the Behavioral Risk Factor Surveillance System.
\item Probability, and a simulation to test whether or not Kobe Bryant has a ``hot hand", i.e. whether he's more likely to hit a shot after a sequence of hits than after a miss. This is students' first exposure to loops in R.
\item Linear regression, and major league baseball data. 
\item Confidence Intervals and data from the housing market in Ames, Iowa. This lab compares the bootstrap confidence interval with the parametric method. 
\item Inference on proportions and data on atheism in America. This lab involves a randomization-based test. 
\item Multivariate linear regression and student evaluation data from UT Austin. In this lab, students discover that student evaluations are biased against women and minorities, and are correlated to attractiveness, hopefully leading them to question their own subconscious biases.
\item ANOVA - beginning with inference for two means, students build up more advanced hypothesis tests on a dataset consisting of births in North Carolina. En route, students must clean the data.
\end{enumerate}

In 2015, I created two labs based on my experience teaching computer science. These labs taught students how to write conditionals, for loops, while loops, and functions. The value of this is demonstrated by a couple of challenging probability problems for which there is no R built-in. I also created a lab where students learn how to sample, plot, and compute p-values for all the major distributions in the course, including the normal, t, $\chi^2$, and $F$-distributions. They also learn how to made Q-Q plots to check whether data fits a distribution. These labs are available at \cite{applied-stats-webpage}, but I subsequently learned that best practices in statistics discourage teaching programming for the sake of programming \cite{gaise-report} (page 24). As a result, I disassembled these labs and assigned relevant bits as part of the daily R practice in the statistics sequence. Examples of the resulting homework assignments can be found in \cite{white-242-2016} and \cite{stat-modeling-webpage}. For more about teaching via advanced computing in R, see \cite{horton-toolbox}.

The next three weeks contain highly structured labs in RStudio. For example, one reviews how to do what they have been doing in DataCamp in RStudio, via the analysis of a built-in dataset (regarding foot length and foot width of children) that comes with the mosaic package in R. 
Students are led through the analysis step by step, and questions ask students to either interpret R output or type a single line of code. 
Another lab, based on a lab of Amy Wagaman of Amherst College (which itself is based on a lab of Ben Baumer of Smith College), teaches students about Tukey's bulging rule for transformations in linear regression, with an application to a dataset of housing prices in Boston. Another walks students through several ANOVA tests on a dataset related to the sale prices of bulls. If an Applied Statistics course cannot reach ANOVA, then this lab can be swapped with another lab based on the Data Repository (Section \ref{subsec:labs-repository}) or with one of the early labs from Statistical Modeling (described below).

In the final weeks, students have minimally structured labs such as the World Bank lab described in Section \ref{subsec:labs-repository}. Another lab, focused on advanced data visualization and data cleaning, is the military spending lab described in Section \ref{sec:applied-topics}. Students create world maps where countries are shaded by the amount they spend annually on the military. The first map is not useful, because of outliers, so students have to take the log. Certain countries in the data need their names to be modified in order to be seen by the map drawing software. Before country names have been modified, the map contains blanks where those countries should be, so students get to test their geography skills when cleaning the data. Students then create related graphics, to study military spending per capita, or as a fraction of GDP. Students write their lab report aimed at convincing a senator to either decrease or increase spending, and select graphics to support their argument. The importance of advanced visualization techniques is highlighted in \cite{stat-graphics, nolan-visualization}.

During the second half of the semester, students are simultaneously working on their final projects, and often come to me with questions. At the beginning, these are usually questions about cleaning the data and getting it read into R (this always takes longer than expected). These meetings with students usually fit within office hours, and give me a chance to detect any potential issues with what a student is trying to do. I try to point out issues of bias in the data, so that students can discuss these in their final reports, and can either correct the bias or (more likely) limit the scope of their analysis accordingly. If the obstacles a student encounters are too severe, I allow students to switch datasets, but I require them to write a new project proposal when they do so.

In the fall of 2015, several students procrastinated on starting the final project until the last possible moment, and thereby deprived themselves of the chance to meet with me as issues arose. The final projects produced by these students often had severe issues that would have been easy to correct if they had been meeting with me. Consequently, in the fall of 2016, I required students to complete weekly ``check-ins" where they would repeat the exact analysis from the weekly lab on their own dataset. I hoped this would keep them making steady progress on their final projects, and would give them lots of material to draw from when the time came to write the final paper. While the resulting final papers were significantly better than in the previous year, students found it burdensome to have both weekly labs and weekly check-ins. In the future, I will make the ``check-in" part of the project, and will shorten the weekly labs slightly to give students time to complete the required tasks for their final projects.

It is an unavoidable fact that meeting with students about their final projects in the second half of the semester takes a lot of time. For me, the resulting final papers are well-worth the investment. Furthermore, taking the time to meet with students makes the grading of the final projects very easy. I can tell from student final presentations if they carried out the plan from their proposal, and if they followed the advice I gave in our meetings. I can then simply skim through the final papers to check that all the details are there and that the writing style is appropriate. The most time-consuming part of helping students with their final projects is learning about datasets I have never seen before, e.g. helping students clean datasets they found online. As course enrollments continue to rise, one strategy to reduce the amount of time spent on this task is to employ teaching assistants. These teaching assistants could be students who had previously taken Applied Statistics, and so had the skill set required to detect bias in data, to clean data, and to get it read into R. Furthermore, the teaching assistants could even receive course credit for this task, instead of money, e.g. if they were enrolled in a directed study called ``Statistical Consulting." \footnote{This idea is the result of a conversation with Lisa Dierker, of Wesleyan University.} Another way to reduce instructor overhead is to allow students to work in groups. Teamwork skills are also an inherently valuable course learning outcome, that will serve students well in whatever they do after the course ends.


\subsection{Teaching Methodology}

In this section, I discuss the methodology that allowed me to cover the applied topics listed above without leaving the students behind, sacrificing core content, or turning statistics into a black box. Having a book that the students could learn from was crucial, and Kaplan's book \cite{kaplan} served this purpose very well. I assigned Kaplan's reading questions for each chapter, and I gave the students access to videos describing the main topics and demonstrating how to conduct analyses in R. I also gave students access to applets that I used in class to emphasize conceptual points. These resources can be found on the course webpage \cite{white-242-2016}.

When students arrived, I answered questions related to the reading for as much of the class period as was required. I then worked through a computer example using RMarkdown (i.e. where students could both see the R code and words typed around the blocks of code). These examples were often drawn from my conversations with colleagues in other departments. Before running each block of code, I solicited student answers about what to do next, how to interpret the output, etc. Students became proficient in translating back and forth between the mathematics and whatever data situation was at hand for the day. Students then worked in groups on another example with scaffolded R code in RMarkdown. 

Each week, I gave students videos showing them how to do things in R that they would need during the week. I used class time to demonstrate concepts (e.g. the sampling distribution) and hypothesis tests visually using applets, then gave students time to play with the applets, vary parameters, etc. I found several repositories of applets online, and link to them from the Applied Statistics webpages \cite{applied-stats-webpage, white-242-2016}. 

When I felt the reading was especially clear and the R code was especially close to either the daily homework or to something the class had done before, then I devoted a day to theory, i.e. the mathematics behind the models. For example, early in the course I showed the class how to derive the simple linear regression equations using partial derivatives. Later, I gave them a worksheet getting the same equations using linear algebra and no calculus (instead, minimization is obtained via the observation that the shortest distance between a point and a line is the perpendicular distance; see \cite{cobb-2nd-mathy}). This treatment is also contained in \cite{kaplan}. 
 Most of our students will go on to be practitioners of statistics, rather than statisticians, so I felt the most important use of class time was to train them to choose, build, and test models, to be cautious and skeptical, and to always check the conditions. I had to keep reminding myself to de-emphasize theory, and in future offerings I plan to do even less. This is probably the hardest part about teaching applied statistics as a mathematician.

\section{Further details on the Statistical Modeling course}

\subsection{Topics and prerequisites}

As discussed in Section \ref{sec:applied-topics}, there are many options for a second course in statistics. I chose a course focusing on regression, but other options include Bayesian statistics, machine learning, categorical data analysis, time series analysis, or mathematical statistics (where the focus is on deriving formulas for standard errors and proving results such as the Gauss-Markov Theorem). Because this course was the highest level statistics course my university offered, I devoted about 25\% of the semester to sampling from these other options so that students would have a rough sense of what a statistician is expected to know. For this, I provided the students with handouts (and a binder to keep them organized). This approach resulted in students well-trained to create and evaluate statistical models, to deal with real world data, and to write reports translating technical results into everyday language. Many students got interested in the advanced topics that the class touched on, and this led to the seminars I offered (and hopefully to the development of a third course in statistics in the future).

For my Statistical Modeling course, I kept the pre-requisites low, to invite in non-math majors who had taken research methods in another department. Based on conversations with partner disciplines, I knew which research methods courses contained enough mathematical depth to allow as prerequisites. I also knew which ones required students to work with a statistical computing language akin to R, rather than a point-and-click interface. The pre-requisites were: one previous course in statistics (including a research methods course), calculus II (which, at my university, includes a discussion of vectors and matrices), and either introductory computer science or prior exposure to programming in a statistical computing language. One benefit of the liberal arts college environment is that the set of interested students was small enough that I could easily check if a given student was sufficiently prepared for the class. It is important to note that our calculus II class includes basics about matrices and vectors, more like calculus III at other universities. I did require knowledge of these topics, so that I could explain the covariance matrix, and the geometric interpretation of correlation as a dot product. The treatment of vectors in \cite{kaplan} is sufficient, so students familiar with \cite{kaplan} would not need calculus II.

I used the Stat2 book \cite{stat2}, and supplemented it with handouts on more advanced topics including principal component analysis, time series analysis, categorical data analysis ($\chi^2$-tests and Fisher's exact test), randomization-based inference (beyond that which is found in \cite{stat2}), visualization techniques in R, data mining, the general linear model via linear algebra, Poisson and negative binomial regression, the generalized linear model, and how to cope with heteroscedasticity (via weighted regression and via heteroscedasticity-robust standard errors). Each week, students completed a project implementing the ideas of the week via an analysis in R of a real-world dataset. 

This course proceeded along the same lines as Applied Statistics, but ran much more smoothly because students already had experience with R. I used reading questions to make sure students completed the reading before class. I used daily R examples and activities for students, plus periodic quizzes where use of the R code was allowed. The daily R code was in DataCamp, so was graded for completion, via screenshots the students submitted. I used the method of marking with the word ANSWER where the answers were in student labs, to reduce grading time of labs. In addition to the weekly labs, I assigned both a midsemester and final project where students would analyze (different) datasets of their own choosing. Students presented their analyses in class, and this made grading the projects relatively easy, since I could usually determine from the presentation if there were any serious issues with the analysis.

Due to the topics coming from mathematical statistics (especially the covariance matrix and the hat matrix), and machine learning (especially dimension reduction techniques such as principal component analysis), students need some familiarity with vectors and matrices. If my department introduces a third course in statistics, it would be easy to shift these topics to the third course and go into more depth in the second course on topics not requiring linear algebra. The prerequisites for the third course could also be set up so that students would only require a first course in statistics, but not Statistical Modeling. In this way, students from our Applied Statistics course could always take another statistics course immediately after, even if Statistical Modeling was not being offered that year (our electives are always offered at most once every 2 years).

\subsection{Projects}

Since all students in Statistical Modeling had either learned R or have a background in programming, DataCamp was not necessary. Instead, we began immediately with labs in RStudio. Students who had not taken Applied Statistics were given access to the projects from that course and strongly encouraged to work through them on DataCamp. The first two weeks of Statistical Modeling were kept relatively light to allow these students to catch up. I had these students schedule regular meetings with the course TA, and with me, to make sure they were getting up to speed on R while the course was reviewing content from Applied Statistics (specifically, content related to randomization-based inference). I provided students with daily R homework at the beginning of the semester (often taken from DataCamp), along with videos from a number of free sources online demonstrating how to build models, verify model conditions, and carry out tests in R. 

I now describe the projects from Statistical Modeling, omitting those already discussed in Section \ref{subsec:labs-repository}. Detailed prompts for these projects can be found at the course webpage \cite{stat-modeling-webpage} and in the data repository. As discussed in Section \ref{sec:projects}, these projects were adapted from labs of Wagaman and Kuiper, with some projects of my own design. Since designing projects from scratch is difficult, I encourage the reader to similarly adapt existing labs the first time the course is taught, to allow more time to learn the material and how best to explain it to the students. 

\begin{enumerate}
\item Randomization-based inference - students use StatKey (or R if they are already familiar with it) to test whether or not there was gender discrimination at Adelphi University in 2009. They write a report aimed at the Dean, who is deciding whether or not to settle a court case. Students conduct multiple t-tests and create side by side boxplots to support their argument. This dataset has outliers that students are expected to find and make decisions about.
\item Simple linear regression - students determine whether or not there is a relationship between sleep deprivation and visual learning, as well as delayed effects days later from earlier sleep deprivation. Students write their report aimed at a psychologist.
\item Logistic regression - students carry out several logistic regressions, including trying to predict bankruptcy of a company, trying to predict which species of fly one has based on measurements, and trying to predict whether an individual has diabetes. Students must clean the datasets and convert certain entries to NA when those entries are impossible. Students choose an audience for their report: most chose to write to a hedge fund manager about the risk of investing in these distressed companies.
\item Multiple linear regression - students choose a proxy for economic growth of a country, then use data from the World Bank to create a multiple linear regression model to predict this proxy variable. Models are built using backwards selection, stepwise regression, and best subsets. Cross-validation is used. Students take on the role of an economic advisor to an African country, and give advice about what changes might improve the economy, based on their model.
\item One-way ANOVA - students study sales data related to bulls and determine which categorical predictor variables are useful to predict sale price. The data contains impossible values that students must correct. Students write their report to clients interested in buying bulls, telling the clients what traits indicate value. 
\item Two-way ANOVA - students analyze bicycle-share data from Washington D.C., and build a model to predict how many bikes will be used on a given day, based on traits of the day. The data requires a great deal of massaging before the model can be built. Students write their report to the mayor, and also discuss the value of sensor network data like the bike-share data.
\item Post-hoc tests - students use simulation to see why corrections such as Bonferroni and Tukey HSD are used. Students then analyze a cancer dataset, where post-hoc tests determine the type(s) of cancer a patient has. Students must advise doctors, potentially with lives on the line, of whether or not to use a conservative or liberal post-hoc test in this situation.
\item Multivariate logistic regression - students use the dataset of Hermon Bumpus on sparrows, choose their own response variables, a build models.
\item Principal Component Analysis and Time Series Analysis - students use simulation methods to understand the hockey stick controversy, a widely circulated graph that shows a sharp increase in global temperature. This graph was created from real data, but using a faulty methodology. Student simulate data that has no trend, then use the faulty methodology and see that the hockey stick shape emerges, showing a sharp uptick as time increases. Students write their report to a senator, trying to explain what happened with this controversy, and why the controversy cannot be used as evidence either for or against global warming.
\end{enumerate}

\subsection{Teaching Methodology}

I used a very similar teaching methodology to what I did in Applied Statistics. I chose a book (Stat2 \cite{stat2}) that was easy to read, even if it did not go into quite as much depth as I would have liked. Students would read before class and would watch a short video working out the mathematics in a simple example. Students also had daily R practice to apply those skills to a real world data problem (using DataCamp) or to a simulation. This was meant to remind students about R, in case they had forgotten, and to help the students who had never seen R before (i.e. those who had used a different programming language in their previous statistics class). I graded the daily R practice for completion, but doing it before class helped students quickly figure out the R code I gave them and how to adapt it to their particular situation on the weekly project. This allowed me to spend class discussing how to choose which model to use, check the conditions for the model, and interpret output from R. The videos and daily assignments are at \cite{stat-modeling-webpage}.

Like in Applied Statistics, when the reading seemed very clear, I would devote a day to theory. For example, after reminding students about the interpretation of correlation as a dot product, I was able to discuss the linear algebra approach to multivariate regression, the hat matrix, and the general linear model. When the class completed logistic regression, I was able to cover the generalized linear model via the natural question of: ``what if our response variable has more than 2 possible values?" I also discussed the theory behind the new hypothesis tests we were conducting (e.g. tests to determine model utility, tests to determine if the residuals were normally distributed, tests for randomness of the data, tests for heteroscedasticity), but omitted proofs. Instead, I emphasized what should be true under the null hypothesis and how to gather evidence against the null hypothesis (i.e. how to come up with sample statistics).

The lab report guidelines, hosted on the course webpage \cite{stat-modeling-webpage}, explain in detail what was expected of students in terms of checking the conditions for their models, and being skeptical regarding potential bias in how a dataset was collected. Examples can also be found in Sections \ref{subsec:labs-repository}, \ref{subsec:242-projects}, and \ref{subsec:401-projects} above. I included the theory days to model for students what it means to try to understand a concept before using it, to open the black box so they'd be more comfortable using it, and to encourage them to take a mathematical statistics course at some point. However, these days were always the most challenging for students, and I believe it would be better to shift them to a third course in statistics, so that the second course could retain its applied focus throughout. 

The strength of both Applied Statistics and Statistical Modeling was the focus on real world data, written lab reports, and the final project. The best class time was time spent in class digging into data sets with R, cleaning data, and making decisions about modeling. While this is a bit unusual for a mathematician, the booming interest in statistics, the course enrollments (and interest from non-majors), the high quality final projects being produced, and the feedback from evaluations all point to the fact that this approach is effective, challenging, and highly relevant to interacting with our data-driven world.


\begin{thebibliography}{10}

\bibitem{akritas}
M.~Akritas.
\newblock {\em Probability \& Statistics for Engineers and Scientists with R}.
\newblock Pearson, 2015.

\bibitem{gaise-report}
Aliaga, Cobb, Cuff, Garfield, Gould, Lock, Moore, Rossman, Stephenson, Utts,
  Velleman, and Witmer.
\newblock Gaise college report, http://www.amstat.org/education/gaise/.
\newblock 2012 (revised printing).

\bibitem{ASA2} American Statistical Association (ASA). 
Discovery with Data: Leveraging Statistics with Computer Science to Transform Science and Society, \textit{www.amstat.org/policy/pdfs/BigDataStatisticsJune2014.pdf} 

\bibitem{ams-guidelines}
ASA/MAA Joint~Committee on~Undergraduate~Statistics.
\newblock Second courses in applied statistics,
  http://www.amstat.org/education/pdfs/second-course-syllabus.pdf.
\newblock 2016 (Feb. 22).

\bibitem{discovery-projects}
Brad Bailey, Dianna J. Spence, and Robb Sinn. Implementation of Discovery Projects in Statistics. Journal of Statistics Education Volume 21, Number 3 (2013), www.amstat.org/publications/jse/v21n3/bailey.pdf

\bibitem{biehler} R. Biehler. Software for Learning and for Doing Statistics. International Statistical Review, 62, 2, 167-189, 1997.

\bibitem{2nd-design} Natalie J. Blades, G. Bruce Schaalje, \& William F. Christensen (2015) The
Second Course in Statistics: Design and Analysis of Experiments?, The American Statistician,
69:4, 326-333,

\bibitem{R-relative} P. Burns. R Relative to Statistical Packages:
Comment 1 on Technical Report Number 1 (Version 1.0)
Strategically using General Purpose Statistics Packages:
A Look at Stata, SAS and SPSS. UCLA Technical Report Series, Report Number 1, Comment Number 1. 2007. http://www.burns-stat.com/pages/Tutor/R\_relative\_statpack.pdf

\bibitem{pcmi} R. D. De Veaux, M. Agarwal, M. Averett, B. Baumer, A. Bray, T. Bressoud, L. Bryant, L. Cheng, A. Francis, R. Gould, A. Kim, M. Kretchmar, Q. Lu, A. Moskol, D. Nolan, R. Pelayo, S. Raleigh, R. Sethi, M. Sondjaja, N. Tiruviluamala, P. Uhlig, T. Washington, C. Wesley, D. White, P. Ye. Curriculum Guidelines for Undergraduate Programs in Data Science. Annual Review of Statistics and Its Application, 4, 2016.

\bibitem{open-intro} Andrew Bray and Mine \c{C}etinkaya-Rundel. ``Labs for R." Web blog post. OpenIntro. The OpenIntro Project, 26 July, 2016.

\bibitem{cobb-2nd-mathy} G. Cobb. Teaching statistics: Some important tensions. Chilean Journal of Statistics
Vol. 2, No. 1, April 2011, 31-62.

\bibitem{cobb-ptolemaic} G. Cobb. The Introductory Statistics Course: A Ptolemaic Curriculum? Technology Innovations in Statistics Education, 1(1), 2007.

\bibitem{stat2}
G.~Cobb B. Hartlaub J. Legler R. Lock T. Moore A.~Rossman Cannon, A. and
  J.~Witmer.
\newblock {\em Stat2: Building Models for a World of Data}.
\newblock WH Freeman, 2012.

\bibitem{asa-guidelines}
B. Chance, S. Cohen, S. Grimshaw, J. Hardin, T. Hesterberg, R. Hoerl, N. Horton, M. Mallone, R. Nichols,
  and D. Nolan.
\newblock Curriculum guidelines for undergraduate programs in statistical
  science, http://www.amstat.org/education/pdfs/guidelines2014-11-15.pdf.
\newblock 2014.

\bibitem{cupm}
2015 CUPM Curriculum Guide to Majors in the Mathematical Sciences
{\textit Mathematics Association of America (MAA)} \textit{http://www.maa.org/sites/default/files/pdf/CUPM/pdf/CUPMguide\textunderscore print.pdf}



\bibitem{freeman} S. Freeman, S. Eddy, M. McDonough, M. Smith, N. Okoroafor, H. Jordt, M.P. Wenderoth. Active learning increases student performance in
science, engineering, and mathematics. PNAS vol. 111 no. 23, 8410--8415, 2014.

\bibitem{stat-graphics} M. Friendly. The Golden Age of Statistical Graphics. Statistical Science
2008, Vol. 23, No. 4, 502-535

\bibitem{halvorsen-projects} K. Halvorsen and T.L. Moore, Motivating, Monitoring, and Evaluating Student Projects.  Proceedings of the Section on Statistical Education, Joint Statistics Meetings of the American Statistical Association, 1991.

\bibitem{hesterberg-bootstrap} T.C. Hesterberg, What Teachers Should Know About the Bootstrap: Resampling in the Undergraduate Statistics Curriculum. Am Stat. 69(4):371-386. 2015.

\bibitem{horton-toolbox} N. J. Horton, E. R. Brown, and L. Qian. Use of R as a Toolbox for Mathematical Statistics Exploration. The American Statistician, 58 (4), 2004.

\bibitem{kaplan}
D.~Kaplan.
\newblock {\em Statistical Modeling: A fresh approach}.
\newblock Ingram, 2009.

\bibitem{taxonomy-dirty} W. Kim, B-Y Choi, E-K Hong, S-K Kim, D Lee. A Taxonomy of Dirty Data. Data Mining and Knowledge Discovery, 7, 81–99, 2003


\bibitem{kuiper-book}
S.~Kuiper and J.~Sklar.
\newblock {\em Practicing Statistics: Guided Investigations for the Second
  Course}.
\newblock Pearson, 2012.

\bibitem{capstone} N.A. Lazar, J. Reeves, and C. Franklin. A Capstone Course for Undergraduate Statistics Majors. The American Statistician, Vol. 65, No. 3 (2011), pp. 183-189

\bibitem{lock-primus}
R.~Lock and P.~Lock.
\newblock Introducing statistical inference to biology students through
  bootstrapping and randomization.
\newblock 18:39--48, 2008.

\bibitem{lock5}
R.~Lock P. Lock K. Lock E. Lock~D. Lock.
\newblock {\em Unlocking the power of statistics}.
\newblock Wiley, 2012.

\bibitem{trevor-R-code}
Trevor Masters.
\newblock Exhaustive power transform, available electronically from
  https://github.com/chukrum47/exhaustivepowertransform.
\newblock 2016.

\bibitem{trevor-genome}
Trevor Masters.
\newblock Distinguishing mental disorders via their genomic roots, preprint available
  electronically from http://personal.denison.edu/~whiteda/research.html.
\newblock 2016.

\bibitem{R-versus} M. Mitchell. Strategically using General Purpose Statistics Packages:
A Look at Stata, SAS and SPSS. UCLA Technical Report Series, Report Number 1, Version Number 1, 2007. http://citeseerx.ist.psu.edu/viewdoc/download?doi=10.1.1.90.1292\&rep=rep1\&type=pdf

\bibitem{moore-math-stat} David S. Moore. Should Mathematicians Teach Statistics? The College Mathematics Journal, Vol. 19, No. 1. (1988), pp. 3-7.

\bibitem{stats-lib-arts} T.L. Moore and R. A. Roberts. Statistics at Liberal Arts Colleges. The American Statistician, Vol. 43, No. 2 (May, 1989), pp. 80-85.

\bibitem{stats-lib-arts2} T.L. Moore and J. A. Witmer. Statistics Within Departments of Mathematics at Liberal Arts Colleges. The American Mathematical Monthly, Vol. 98, No. 5 (1991), pp. 431-436.

\bibitem{nolan-explorations} Deborah Nolan and Duncan Temple Lang. Explorations in Statistics Research:
An Approach to Expose Undergraduates to
Authentic Data Analysis. The American Statistician 69, 292-299, 2015.

\bibitem{nolan-visualization} Deborah Nolan and Jamis Perrett. Teaching and Learning Data Visualization: Ideas
and Assignments. The American Statistician 70, 260-269, 2016.

\bibitem{nolan-computing-curr} Deborah Nolan and Duncan Temple Lang (2010) Computing in the Statistics Curricula, The American Statistician, 64:2, 97-107, DOI: 10.1198/tast.2010.09132

\bibitem{nolan-speed} Deborah Nolan and Terry Speed.
Teaching Statistics Theory Through Applications. The American Statistician 53, 370-375, 1999.

\bibitem{nolan-book} Deborah Nolan and Terry Speed. {\em Stat Labs: Mathematical Statistics Through Applications}. New York: Springer-Verlag, 2000.

\bibitem{beyond-normal} B. Smucker and A. J. Bailer. Beyond Normal: Preparing Undergraduates for the Work Force in a Statistical Consulting Capstone.  The American Statistician, Volume 69, 300-306, 2015 

\bibitem{reproducibility} V. Stodden. Reproducing Statistical Results. Annu. Rev. Stat. Appl. 2015. 2:1-19.

\bibitem{combating-simulation} N. Tintle, B. Chance, G. Cobb, S. Roy, T. Swanson, J. VanderStoep. Combating anti-statistical thinking using simulation-based methods throughout the undergraduate curriculum. Faculty Work: Comprehensive List. Paper 574, 2015. http://digitalcollections.dordt.edu/faculty\_work/574

\bibitem{tintle-rand-based} N. Tintle, J. VanderStoep, V-L Holmes, B. Quisenberry, T. Swanson. Development and assessment of a preliminary randomization-based introductory statistics curriculum. Journal of Statistics Education Volume 19, Number 1 (2011).

\bibitem{tybl}
Alexander~J. Tybl.
\newblock An overview of spatial econometrics, preprint available
  electronically from https://arxiv.org/abs/1605.03486.
\newblock 2016.

\bibitem{spurious} Tyler Vigen. Spurious Correlation website: \href{http://www.tylervigen.com/spurious-correlations}{http://www.tylervigen.com/spurious-correlations}, accessed September, 2016.

\bibitem{wagaman} A. Wagaman. Meeting Student Needs for Multivariate Data Analysis: A Case Study in Teaching a Multivariate Data Analysis Course with No Pre-requisites. The American Statistician 70, 405-412, 2016.

\bibitem{statcrunch} W. West. Social Data Analysis with StatCrunch: Potential Benefits to Statistical Education. Technology Innovations in Statistics Education, 3(1), 2009.

\bibitem{webpage}
David White.
University Webpage: \href{http://personal.denison.edu/~whiteda/index}{http://personal.denison.edu/$\sim$whiteda/index}.

\bibitem{applied-stats-webpage}
David White.
\newblock Math 242: Applied statistics, course webpage,
  \href{http://personal.denison.edu/~whiteda/math242fall2015.html}{http://personal.denison.edu/$\sim$whiteda/math242fall2015.html},
  2015.
  
  \bibitem{white-242-2016}
David White.
\newblock Math 242: Applied statistics, course webpage,
  \href{http://personal.denison.edu/~whiteda/math242fall2016.html}{http://personal.denison.edu/$\sim$whiteda/math242fall2016.html},
  2016.

\bibitem{stat-modeling-webpage}
David White.
\newblock Math 401: Statistical modeling, course webpage,
  \href{http://personal.denison.edu/~whiteda/math401spring2016.html}{http://personal.denison.edu/$\sim$whiteda/math401spring2016.html},
  2016.
  
  \bibitem{Wild and Pfannkuch}
C.J. Wild and M. Pfannkuch, Statistical Thinking in Empirical Enquiry \textit{International Statistical Review}, 67(3) 223-265. 1999.

\bibitem{accessible-inference} C.J. Wild, M. Pfannkuch, and M. Regan. Towards more accessible conceptions of statistical
inference. Journal of the Royal Statistical Society, 174 (2), 247-295, 2011.
  
  \bibitem{accessible-bootstrap}  C.J. Wild, M. Pfannkuch, M. Regan, and R. Parsonage. Accessible Conceptions of Statistical
Inference: Pulling Ourselves Up by
the Bootstraps. International Statistical Review 85, 1, 84-107, 2017.
  
  \bibitem{isostat}
  Jeffrey Witmer. Isostat email listserv, \href{http://ww2.amstat.org/committees/isostat/isostat.html}{http://ww2.amstat.org/committees/isostat/isostat.html}.
  
\bibitem{wood} M. Wood.   The Role of Simulation Approaches in Statistics. Journal of Statistics Education Volume 13, Number 3 (2005), ww2.amstat.org/publications/jse/v13n3/wood.html
  
  \bibitem{wooldridge} J. Wooldridge, {\em Introductory Econometrics: a Modern Approach}. Mason, Ohio: South-Western Cengage Learning, 1960 (2012).
  
  \bibitem{preprocessing} Y. Zhu, L.M. Hernandez, P. Mueller, Y. Dong, M.R. Forman. Data Acquisition and Preprocessing in Studies on Humans:
What Is Not Taught in Statistics Classes?, American Statistician 67(4), 235-241, 2013.

\end{thebibliography}
\end{document}